\input harvmac
\def\lf{16\pi^2}
 
\def\ga{\gamma}

\def\frak#1#2{{\textstyle{{#1}\over{#2}}}}

\def\sy{supersymmetry}
\def\sic{supersymmetric}
   
\def\ssm{supersymmetric standard model}

\def\pa{\partial}

\def\npb{{Nucl.\ Phys.\ }{\bf B}}

\def\plb{{Phys.\ Lett.\ }{\bf B}}

\def\llf{(16\pi^2)^2}
\def\lllf{(16\pi^2)^3}

{\nopagenumbers
\line{\hfil LTH 375}
\line{\hfil hep-ph/9606323}
\vskip .5in
\centerline{\titlefont $N=1$ supersymmetry and the 
three loop}
\centerline{\titlefont gauge $\beta$-function}
\vskip 1in
\centerline{\bf I.~Jack, D.R.T.~Jones and C.G.~North}
\bigskip
\centerline{\it Department of Mathematical Sciences, 
University of Liverpool, Liverpool L69 3BX, U.K.}
\vskip .3in

We calculate the three loop gauge $\beta$-function for an abelian 
$N=1$ supersymmetric gauge theory, using DRED. We construct 
a coupling constant redefinition that relates the result to the 
corresponding term in the NSVZ $\beta$-function, and by generalising 
this redefinition to the non-abelian case we derive the DRED 
three loop gauge $\beta$-function for the non-abelian case.

\Date{June 1996}}

In a previous paper \ref\jjn{I.~Jack, D.R.T.~Jones and C.G.~North,
hep-ph/9603386} we calculated the 
 three-loop  contribution $\ga^{(3)}$  to the anomalous 
dimension of the chiral supermultiplet  in a general  $N=1$
supersymmetric gauge theory, using DRED.
\footnote{\dag}{By DRED we mean dimensional reduction
\ref\siegel{W.~Siegel, \plb 84B (1979) 193\semi
D.M.~Capper, D.R.T.~Jones and P.~van Nieuwenhuizen,
\npb 167 (1980) 479}\ with minimal (or modified minimal) subtraction.}
 Here we present the analogous result for the 
gauge $\beta$-function. In fact we perform the explicit calculation 
only for the abelian case, and infer the non-abelian result by comparing 
our result to the all-orders NSVZ $\beta$-function\ref\nov{V.A.~Novikov et al,
\plb166 (1986) 329\semi 
M.A.~Shifman and A.I~Vainstein, \npb 277 (1986) 456}.

The Lagrangian $L_{\rm SUSY} (W)$ for a $N=1$ \sic\ theory 
is defined by the superpotential
\eqn\Ea{
W={1\over6}Y^{ijk}\Phi_i\Phi_j\Phi_k+{1\over2}\mu^{ij}\Phi_i\Phi_j. }
$L_{\rm SUSY}$ is the Lagrangian for  the $N=1$ supersymmetric
gauge theory, containing the gauge multiplet 
and a chiral superfield $\Phi_i$ with component fields
$\{\phi_i,\psi_i\}$ transforming as a 
representation $R$ of the gauge group $\cal G$. 
We assume that there are no gauge-singlet
fields and that $\cal G$ is simple.

The $\beta$-functions for the Yukawa couplings $\beta_Y^{ijk}$ 
are given by
\eqn\Ec{
\beta_Y^{ijk}= Y^{p(ij}\ga^{k)}{}_p = 
Y^{ijp}\ga^k{}_p+(k\leftrightarrow i)+(k\leftrightarrow j),}
where $\ga$ is the anomalous dimension for $\Phi$. 
The one-loop results for the gauge coupling $\beta$-function $\beta_g$ and 
for $\ga$ are given by
\eqn\Ed{
\lf\beta_g^{(1)}=g^3Q,\quad\hbox{and}\quad 
\lf\ga^{(1)i}{}_j=P^i{}_j,}
where
\eqna\Ee$$\eqalignno{ 
Q&=T(R)-3C(G),\quad\hbox{and}\quad &\Ee a\cr
P^i{}_j&={1\over2}Y^{ikl}Y_{jkl}-2g^2C(R)^i{}_j. &\Ee b\cr}$$
Here
\eqn\Ef{
T(R)\delta_{AB} = \Tr(R_A R_B),\quad C(G)\delta_{AB} = f_{ACD}f_{BCD} 
\quad\hbox{and}\quad C(R)^i{}_j = (R_A R_A)^i{}_j.}

The two-loop
$\beta$-functions for the dimensionless couplings were calculated in
\break Refs.~\ref\tja{D.R.T.~Jones, \npb87 (1975) 127}%
\nref\pwa{A.J.~Parkes and P.C.~West, \plb138 (1984) 99}%
\nref\pwb{A.J.~Parkes and P.C.~West, \npb256 (1985) 340}%
\nref\west{P.~West, \plb137 (1984) 371}%
--\ref\tjlm{D.R.T. Jones and L. Mezincescu, \plb136 (1984) 242; 
{\it ibid} 138 (1984) 293}: 
\eqna\Au$$\eqalignno{ \llf\beta_g^{(2)}&=2g^5C(G)Q-2g^3r^{-1}C(R)^i{}_jP^j{}_i
&\Au a\cr
\llf\ga^{(2)i}{}_j&=[-Y_{jmn}Y^{mpi}-2g^2C(R)^p{}_j\delta^i{}_n]P^n{}_p+
2g^4C(R)^i{}_jQ,&\Au b\cr}
$$
where $Q$ and $P^i{}_j$ are given by Eq.~\Ee{}, and $r=\delta_{AA}$. 

In our notation the NSVZ formula for $\beta_g$ is 
\eqn\russa{\beta_g^{NSVZ} = 
{{g^3}\over{\lf}}\left[ {{Q- 2r^{-1}\Tr\left[\ga C(R)\right]}
\over{1- 2C(G)g^2{(\lf)}^{-1}}}\right],}
which leads to 
\eqn\russab{\eqalign{\lllf\beta^{(3)NSVZ}_g =& 
4g^7 Q C(G)^2 -4g^5 C(G) r^{-1} \lf \Tr\left[ \ga^{(1)}C (R)\right] 
\cr&-2g^3 r^{-1} \llf \Tr\left[ \ga^{(2)}C (R)\right].\cr}}

Note that $\beta^{(3)NSVZ}_g$ vanishes for a one-loop finite theory. 
This holds also for $\beta^{(3)DRED}_g$, as  explicitly verified 
by Parkes and West\pwb; see also 
Ref.~\ref\gmz{M.T.~Grisaru, B.~Milewski and D.~Zanon, \plb 155 (1985)357}.
We also have for a theory with $N=2$ \sy\ that 
$\beta^{(3)NSVZ}_g = 0$, as is easy to verify from 
Eq.~\russab, and $\beta^{(3)DRED}_g = 0$,  because 
$N=2$ theories have one loop divergences only\ref\hsw{
P.S.~Howe, K.S.~Stelle and P.~West, \plb 124 (1983) 55\semi
P.S.~Howe, K.S.~Stelle and P.K.~Townsend, \npb  236 (1984) 125}.
Nevertheless we shall see that DRED does not give the NSVZ result at three 
loops. 

Let us turn to the explicit calculation. In the abelian case, this amounts 
to a straightforward determination using standard Feynman rules
\ref\grsb{M.~Grisaru, W.~Siegel and M.~Rocek, \npb 159 (1979) 429} 
of the vector supermultiplet 
self--energy. Now in the special case of one-loop finite theories, 
Parkes and West\pwb\ were able to derive the result $\beta^{(3)DRED}_g = 0$    
in the non-abelian case via an essentially abelian calculation, 
by using the fact that the same result holds for $N=2$ theories 
(whether one-loop finite or not). 
As we shall see, this property of 
$N=2$ theories will be useful for us as well. 

Our result is 
\eqn\russac{\eqalign{
\lllf\beta^{(3)DRED}_g &= 3r^{-1}g^3Y^{ikm}Y_{jkn}P^n{}_mC(R)^j{}_i
+6r^{-1}g^5\tr\left[PC(R)^2\right]\cr&+r^{-1}g^3\tr\left[P^2C(R)\right]
-6r^{-1}Qg^7\tr\left[C(R)^2\right]}}
where here $Q=T(R)$. 
Now we compare this with the corresponding result for $\beta^{(3)NSVZ}_g$, 
obtained by setting $C(G) = 0$ in Eq.~\russab. They are not the same, 
but if we redefine the DRED coupling $g$ as follows:

\eqn\tlfba{
g \to g + \delta g, \quad\hbox{where}\quad 
\delta g = -(\lf)^{-2}\frak{1}{2r}g^3\tr \left[PC(R)\right] 
}
then the resulting change in $\beta_g$, $\delta\beta_g$ satisfies
\eqn\tlfe{\eqalign{\delta\beta_g &= \left[\beta_Y. {{\pa}\over{\pa Y}} 
+ \beta_Y^* . {{\pa}\over{\pa Y^*}} 
  + \beta_g . {{\pa}\over{\pa g}}\right]\delta g
- \delta g . {{\pa}\over{\pa g}} \beta_g\cr
&= -r^{-1}g^3Y^{ikm}Y_{jkn}P^n{}_mC(R)^j{}_i
-2r^{-1}g^5\tr\left[PC(R)^2\right]\cr&-r^{-1}g^3\tr\left[P^2C(R)\right]
+2r^{-1}Qg^7\tr\left[C(R)^2\right],\cr}}
and it is easy to show that
\eqn\russad{\beta^{(3)NSVZ}_g = \beta^{(3)DRED}_g + \delta\beta_g.}
Notice that it is quite non-trivial that 
$\beta^{(3)NSVZ}_g$  and $\beta^{(3)DRED}_g$ can be related in this way; 
$\delta g$  as defined in Eq.~\tlfba\ leads to four distinct tensor 
structures in Eq.~\tlfe.

We turn now to the non-abelian case. The crucial observation is that 
$\delta g$ as defined in Eq.~\tlfba\ {\it does not vanish\/} for 
a $N=2$ theory in general (though it does in the abelian case, 
as may be easily verified). There is, however, an obvious generalisation 
of it to the non-abelian case, to wit 
\eqn\tlfbb{
\delta g = (\lf)^{-2}\frak{1}{2}g^3 
\left[r^{-1}\tr\left[PC(R)\right]-g^2QC(G)\right] 
}
where we have reversed the overall sign (compared to Eq.~\tlfba) 
because we plan to use this $\delta g$ to go back from 
$\beta^{(3)NSVZ}_g$ to $ \beta^{(3)DRED}_g$. It is easy to verify that 
Eq.~\tlfbb\ leads to $\delta g = 0$ in the $N=2$ case.  
Are there any other candidate terms for inclusion in $\delta g$? 
We are constrained by the following requirements:
\item{(1)}
$\delta g = 0$ for a one-loop finite theory.
\item{(2)}
$\delta g = 0$ for a $N=2$ theory. 
\item{(3)}
Eq.~\tlfba\ must hold in the abelian case.
\item{(4)}
The resulting terms in $\delta\beta_g$ must correspond to possible 
1PI Feynman graphs.

It is easy to convince oneself that Eq.~\tlfbb\ represents the only 
possible transformation (up to an overall constant, which we have 
fixed by the abelian calculation). With hindsight, in fact, simply calculating 
the coefficient of the $\tr\left[P^2 C(R)\right]$ term in $\beta^{(3)DRED}_g$
would have sufficed: much easier than the full abelian calculation.  By 
performing this we have, however, verified that the NSVZ $\beta$-function 
corresponds to a scheme equivalent to DRED. 

Our result for  $\beta^{(3)DRED}_g$ in the non-abelian case 
is therefore:

\eqn\russaf{\eqalign{
\lllf\beta^{(3)DRED}_g &= 3r^{-1}g^3Y^{ikm}Y_{jkn}P^n{}_mC(R)^j{}_i
+6r^{-1}g^5\tr\left[PC(R)^2\right]\cr&+r^{-1}g^3\tr\left[P^2C(R)\right]
-6r^{-1}Qg^7\tr\left[C(R)^2\right] - 4r^{-1}g^5C(G)\tr\left[PC(R)\right]\cr&
+g^7QC(G)\left[4C(G)-Q\right].\cr}}  

Of course our method has been somewhat indirect so it would be 
interesting to confirm Eq.~\russaf\ by an explicit calculation. 
Remarkably enough, in the special case of no chiral  superfields, there 
does exist one in the literature, by 
Avdeev and Tarasov\ref\avdeev{L.V~Avdeev and O.V.~Tarasov \plb 112 (1982) 356}. 
They obtained the result 
\eqn\avtar{
\beta^{DRED}_g = -3g^3C(G)(\lf)^{-1}-6g^5C(G)^2(\lf)^{-2}
-21g^7C(G)^3(\lf)^{-3}+\cdots}
which precisely agrees with Eq.~\russaf. Now Ref.~\avdeev\ employed 
DRED with component fields rather than superfields, and hence a very 
different (and not manifestly \sic) gauge; as we should 
perhaps have expected, however, 
within DRED $\beta_g$ is gauge invariant. 
(For discussion of the gauge invariance of $\beta_g$ in the 
context of ordinary dimensional regularisation, see 
Ref.~\ref\jcw{D.R.T.~Jones, \npb 75 (1974)\semi
W.E.~Caswell and F.~Wilczek, \npb 49 (1974) 291}). 
Thus our conjecture in Ref.~\jjn\ that DRED would 
reproduce $\beta^{(3)NSVZ}_g$ was perhaps misguided. 

In conclusion: our main result is the DRED result for  the three loop
gauge $\beta$-function, Eq.~\russaf. In Ref.~\ref\fjj{P.M.~Ferreira,
I.~Jack and D.R.T.~Jones, hep-ph/9605440} the results of Ref.~\jjn\ and
this paper are used to derive the  three-loop \ssm\ $\beta$-functions,
and investigate their effect on the standard running coupling
analysis. (For an interesting alternative approach  to this running
analysis, see Ref.~\ref\shif{M.~Shifman, hep-ph/9606281}).

\bigskip\centerline{{\bf Acknowledgements}}\nobreak

IJ and CGN were supported by PPARC via an Advanced Fellowship and a 
Graduate Studentship respectively. 

\listrefs 

\bye